\documentclass[conference]{IEEEtran}
\usepackage{cite}
\usepackage{amsmath,amssymb,amsfonts}
\usepackage{algorithmic}
\usepackage{graphicx}
\usepackage{textcomp}
\usepackage{xcolor}
\usepackage{threeparttable}
\usepackage{multirow}
\usepackage{placeins}
\usepackage[colorlinks=true,citecolor=blue]{hyperref}
\def\BibTeX{{\rm B\kern-.05em{\sc i\kern-.025em b}\kern-.08em
    T\kern-.1667em\lower.7ex\hbox{E}\kern-.125emX}}
\begin{document}

\title{The Paradox of Spreadsheet Self-Efficacy: Social Incentives for Informal Knowledge Sharing in End-User Programming\\
\thanks{This work is supported by the Engineering and Physical Sciences Research Council [Grant number: EP/W522077/1] and Microsoft Research through its EMEA PhD Scholarship Programme}
}

\author{\IEEEauthorblockN{Qing (Nancy) Xia\textsuperscript{1}, Advait Sarkar\textsuperscript{1, 2, 3}, Duncan P. Brumby\textsuperscript{1}, Anna Cox\textsuperscript{1}}
\IEEEauthorblockA{
\textit{\textsuperscript{1}University College London, \textsuperscript{2}Microsoft Research, \textsuperscript{3}University of Cambridge}, United Kingdom\\
nancy.xia.18@ucl.ac.uk, advait@microsoft.com, d.brumby@ucl.ac.uk, anna.cox@ucl.ac.uk}
}

\maketitle

\IEEEpeerreviewmaketitle 

\vspace{-1cm}
\begin{abstract}
Informal Knowledge Sharing (KS) is vital for end-user programmers to gain expertise. To better understand how personal (self-efficacy), social (reputational gains, trust between colleagues), and software-related (codification effort) variables influence spreadsheet KS intention, we conducted a multiple regressions analysis based on survey data from spreadsheet users (\textit{n}=100) in administrative and finance roles. We found that high levels of spreadsheet self-efficacy and a perception that sharing would result in reputational gains predicted higher KS intention, but individuals who found knowledge codification effortful showed lower KS intention. We also observed that regardless of occupation, users tended to report a lower sense of self-efficacy in their general spreadsheet proficiency, despite also reporting high self-efficacy in spreadsheet use for job-related contexts. Our findings suggest that acknowledging and designing for these social and personal variables can help avoid situations where experienced individuals refrain unnecessarily from sharing, with implications for spreadsheet design.\footnote[0]{This is the accepted version of a paper accepted for presentation at the IEEE Symposium on Visual Languages and Human-Centric Computing (VL/HCC), 2024.}

\end{abstract}

\begin{IEEEkeywords}
spreadsheet, knowledge sharing, regression analysis, self-efficacy, codification effort
\end{IEEEkeywords}

\section{Introduction}
End-user programmers, such as spreadsheet users, write code for their own use, but typically have no formal training in computing or programming \cite{ko2011}. In this paper, we are interested in a particularly common and effective method of expertise acquisition among end-user programmers in the workplace: \emph{Knowledge Sharing} (KS) \cite{kiani2020}. Previous research highlights that KS activities, such as interactions in public forums \cite{hung2015}, documentation creation practices \cite{horvath2023, srinivasaragavan2021}, code-sharing, and informal over-the-shoulder recommendations \cite{kiani2020} provide valuable opportunities for user learning. Such interactions are particularly important in the context of spreadsheets, as users often lack formalised training support from organisations \cite{lawson2009}, and are thus reliant on the serendipitous discovery of knowledge from these sharing interactions to develop their proficiency \cite{sarkar2018}. Moreover, important contextual and technical knowledge embedded in spreadsheets is usually accessible only to the author \cite{srinivasaragavan2021, kohlhase2015}, so without proper sharing practices, this knowledge can easily be lost \cite{smith2017}.

Given the critical role of KS in enhancing spreadsheet users' proficiency and learning, understanding the factors which influence spreadsheet knowledge sharing intention is important for designing tools which can facilitate engagement with sharing activities. A common approach for tools developed to facilitate KS is to reduce the effort of interacting with the software in the process of sharing through automation (e.g. automatically extracting and highlighting users' feature usage patterns on the interface \cite{matejka2009}). However, designs aimed at facilitating KS may encounter challenges in user adoption and effectiveness if they fail to adequately consider the social and personal influences which motivate sharing practices \cite{giannisakis2022, pipek2003}. One characteristic of the sharer which is particularly relevant for spreadsheet users is self-efficacy -- an individual's belief in their own ability to successfully achieve a positive outcome (e.g. their ability to provide valuable contributions in KS, or ability to use spreadsheets well) \cite{bandura2001}. Low spreadsheet self-efficacy, reflected in a lack of faith in one's spreadsheet expertise, could cause users to refrain unnecessarily from KS. 

The aim of this paper is to identify the factors which affect spreadsheet KS intention and clarify the relationships. Understanding these factors can inform the design and implementation of technical solutions to facilitate spreadsheet-related KS, and to address the risk of long-term knowledge loss \cite{smith2017}. We pose the following research questions. \textit{RQ1: To what extent do personal, social, and software-related constructs influence spreadsheet KS intention?} \textit{RQ2: Are there differences in the way spreadsheet-related self-efficacy is evaluated by users of differing levels of expertise?} 

The main contribution of our paper is a quantitative analysis measuring the extent to which personal, social, and software-related variables influence spreadsheet KS, as well as potential design recommendations for future KS systems. 

\section{Related works and hypotheses}
To identify the factors relevant for our survey, we first reviewed existing research of spreadsheet sharing, recommendation, annotation, and general collaborative activities \cite{sarkar2018, srinivasaragavan2021, chalhoub2022} to identify factors relevant to spreadsheet KS. We then conducted a literature review of papers which applied regression or Structural Equation Modelling approaches to model general KS, and to identify common factors and scales used in measurements. This led to us reviewing leading journals in the field of knowledge management \cite{serenko2009}, information systems, as well as human-computer interaction (e.g. \cite{nguyen2020, kankanhalli2005, lin2009, hsu2007}.) 

Five predictor variables (one personal variable, three social variable, and one software variable) were then identified from these KS models which: a) mapped to key themes in the spreadsheet context; b) had important implications for system designs (which led to the exclusion of organisation-related factors, such as financial incentives); c) had no conceptual overlap between them, as the predictor variables in a regression analysis should not be correlated with each other. Below, we outline the variables and corresponding hypotheses. 

\subsection{Personal variable - Self-efficacy}
Defined as the belief in one’s ability to achieve desired results \cite{bandura2001}, self-efficacy enhances confidence in the value of one’s knowledge and the positive outcomes of sharing \cite{lin2007, kankanhalli2005, nguyen2020, wasko2000}. We consider spreadsheet self-efficacy as an individual's confidence in their spreadsheet proficiency, where proficiency reflects one's spreadsheet knowledge and ability to complete spreadsheet-related tasks. Individuals with high spreadsheet self-efficacy are therefore potentially more likely to engage in KS due to greater confidence in one's contribution in the process. However, users expressing low self-efficacy may also be unintentionally downplaying their proficiency. Previous research highlights that technical spreadsheet proficiency is often expressed in the context of one's domain knowledge \cite{kankuzi2016, chambers2010}. Therefore, it is important to consider not only general spreadsheet self-efficacy (GSE), which concerns general technical spreadsheet proficiency, but also context-specific self-efficacy (CSSE), or proficiency in the context of one's work-related tasks. We hypothesise that:

\textit{H1a: Self-efficacy positively affects spreadsheet knowledge sharing intention}

\textit{H1b: There is a difference in self-reported ratings for statements of general spreadsheet self-efficacy compared to statements of context-specific spreadsheet self-efficacy.}

\subsection{Social variables - Reputation and trust}
Previous work suggests that the propagation of spreadsheet knowledge is often driven through organisations by select individuals, who build reputations as experts through KS \cite{nardi1991, nardi1999b}. These experts may even be formally recognised and rewarded for their KS work \cite{nardi1999b}, though some studies suggest the incentive of reputational gains alone may be even more powerful compared to monetary rewards \cite{choi2008, kankanhalli2005}. Considering the informal and serendipitous nature of most spreadsheet-based learning \cite{sarkar2018}, acquiring social reputations may be a key motivator to spreadsheet KS. 

\textit{H2: Reputational gains positively affects spreadsheet knowledge sharing intention.}

Trust serves to both motivate and maintain continuous engagement in KS activities \cite{wickramasinghe2012, lee2010}, and is essential for establishing long-term pro-sharing norms \cite{kankanhalli2005}. In spreadsheet collaboration, the process of working with others is often fraught with errors and misinterpretations for users \cite{srinivasaragavan2021, panko2016, caulkins2007}. Thus, beliefs of how receptive another individual might be to potential recommendations or criticisms (i.e. \textit{disclosure-based trust}), and their belief in another users' technical competence and efficiency of uptake in learning (i.e. \textit{reliance-based trust}) is likely to play a significant role in deciding whether users feel comfortable enough to impart knowledge in common collaboration processes \cite{gillespie2015}. We therefore suggest that: 

\textit{H3: Reliance-based trust in colleagues positively affects spreadsheet knowledge sharing intention.}

\textit{H4: Disclosure-based trust in colleagues positively affects spreadsheet knowledge sharing intention.}

\subsection{Software variable - Codification effort}
Documenting and sharing tacit knowledge is time-consuming and cognitively intensive \cite{kankanhalli2005, davenport2000}, which can discourage voluntary engagement with KS. While spreadsheet authors aim for clear communication, the layered design of the spreadsheet interface and the scale of the datasets involved often obscures the visibility of key information, thus requiring additional clarification from users \cite{srinivasaragavan2021, kohlhase2015, chalhoub2022}. Reports of high codification effort is expected to reflect a view that knowledge sharing in the spreadsheet context is time-consuming and costly, and thus more likely to associate with lowered knowledge sharing intention. 

\textit{H5: Codification effort negatively affects spreadsheet knowledge sharing intention.}

\section{Method}
\subsection{Participants}
153 participants were initially recruited using the online recruitment platform Prolific. We applied a filter to recruit individuals in either finance- or administration-related job functions. This was to ensure that our sample was fairly distributed in terms of overall spreadsheet knowledge, frequency of use, and task type, as previous surveys suggests that individuals in finance-related jobs tend to be more advanced in these areas compared to those in administration-related jobs \cite{lawson2009}. As we were interested in informal, voluntary knowledge sharing interactions, we excluded 48 participants as they had official teaching responsibilities within their organisation. 5 responses were also discarded after quality checks, leaving a total of 100 participants in the remaining analyses. This resulted in 50 finance users and 41 administration users, and 9 others who described their job function as `Other' (see Appendix \ref{sec:appendix_A_participant_demog} for full participant demographic breakdown by occupation). 

\subsection{Measures}
\label{subsec:survey_measures}
Due to the specific interests of this study, we developed our own measure of spreadsheet knowledge sharing intention (KSI). To measure intention in the sense of whether an individual is proactive in engaging in knowledge sharing \cite{davenport2000}, we reviewed the spreadsheet literature to identify common work-related circumstances where KS is not mandatory, but can still have clear benefits for others if an individual chooses to engage in it. We therefore developed a measure to reflect individual willingness to: write documentation \cite{srinivasaragavan2021, chalhoub2022}, share spreadsheet-related resources and tips with others \cite{chambers2010}, give advice \cite{sarkar2018}, and to respond positively to help-seeking interactions \cite{kiani2020}. A follow-up question asked participants to select their most commonly used methods for communicating knowledge in spreadsheets. In line with scales used in knowledge management literature, our scale adopted a broad definition of KS and incorporated both public-facing and one-on-one KS interactions to better capture overall user intention. Both questions were reviewed by two professional spreadsheet users in finance and administration-related jobs respectively and refined based on their feedback. 

For the remaining measures, we used scales from established literature and adapted wording to fit the spreadsheet context. To measure spreadsheet self-efficacy, we referred to the literature review and categorisation of existing software-related self-efficacy scales from Gupta and Bostrom \cite{gupta2019}, as this paper was closest to ours in its position regarding self-efficacy. However, contrary to Gupta and Bostrom, who distinguish between self-efficacy for `complex' and `simple' tasks, we focus on a higher-level distinction between context-specific (CSSE) and general self-efficacy (GSE). Measures of CSSE (items 1-5) are characterised by statements which specifically require participants to answer in the self-imagined context of their job \cite{compeau1995}, while statements of GSE (items 6-9) referred more broadly to one's confidence in their spreadsheet proficiency outside of the job context \cite{gupta2019} (see Appendix \ref{sec:appendix_B_SE_items}).

For the remaining items, we adapted scales from knowledge management and organisational psychology research, with wording changes to fit the spreadsheet context. To measure trust in colleagues, we used the Behavioural Trust Index (BTI) by Gillespie \cite{gillespie2003}, a validated scale specifically designed for professional contexts which captures two factors of trust -- reliance-based trust and disclosure-based trust \cite{gillespie2015}. To measure perceived reputational gains for KS, we referred to well-established survey items in knowledge management \cite{kankanhalli2005, wasko2005}. Finally, the scale for codification effort was taken from \cite{kankanhalli2005}, as the original items also explored user effort with engaging in digital systems (i.e. electronic knowledge repositories), and so could be easily adapted to explore the perceived effort associated with encoding spreadsheet-associated knowledge.

A 7-point Likert scale was used for all questions, with the exception of the questions related to self-reported expertise, which was measured using a 5-point Likert scale. Participants' demographics were collected using items from the Spreadsheet Engineering Research Project (SERP) survey \cite{lawson2009, baker2006}. The final survey consisted of 19 questions (see Appendix \ref{sec:appendix_C_full_survey} for link to the full survey).

\section{Results}

\subsection{Validation}
To ensure the internal consistency of the items developed, we calculated Cronbach's alpha for each of the measured constructs. The recommended acceptable level for Cronbach's alpha is 0.7 or above \cite{nunnally1994}. All scales, with the exception of KSI, scored 0.7 or higher (see Appendix \ref{sec:appendix_D_scale_reliability}), indicating good reliability in the chosen scales. For KSI, Cronbach's alpha fell just below 0.7 to 0.69, but this weakness in internal consistency was attributed to Item 5 only, which had been negatively worded for attention checking purposes, therefore we suspect it was simply the negative wording which may have produced this effect. As this issue manifested for only one item, we believe the reliability of the overall KSI measure was still robust. 

\subsection{Knowledge sharing practices}
We asked participants to report their most commonly used KS method with a multiple choice question. Our results showed that KS activities people most commonly occurred through direct written communications (e.g. emails, Teams channels) and informal conversations, and less commonly through communication with a wider community. Appendix \ref{sec:appendix_E_KS_practices} displays the distribution of responses. 

\subsection{RQ1: Regression analysis}
We conducted a multiple regressions analysis to examine the effects of codification effort, reliance-based trust, disclosure-based trust, self-efficacy, and reputational gains on spreadsheet knowledge sharing intention (Model 1). The overall model offers statistically significant explanatory power for knowledge sharing intention (\textit{F}(5, 94)=6.948, \textit{p}$<$.001, $R^2$=.270). The model accounts for just over a quarter of the variance in spreadsheet KSI, which appears low, and is possibly due to the limited number of factors chosen. Inspection of the variance inflation factors (VIF) in each variable revealed scores ranging from 1.115 to 1.206, where a VIF of 1 indicates no correlation and a VIF greater than 5 indicates high correlation. Our results show that the chosen variables meet the multicollinearity criteria for regression and do not strongly correlate with each other. 

As shown in Table \ref{tab:regression_results}, in Model 1, all variables except for the two trust-related factors predicted knowledge sharing intention with statistical significance, supporting H1a, H2, H5. Based on the model, knowledge sharing intention is negatively related to codification effort -- the lower the perceived effort of codifying one's spreadsheet knowledge, the greater the intention to share one's knowledge (\textit{p}=.003). On the other hand, a greater sense of self-efficacy (\textit{p}=.038) and a greater expectation of reputational gains as a result of sharing (\textit{p}=.027), is positively related to knowledge sharing intention. However, there was no significant relationship between knowledge sharing intention and either reliance-based trust (\textit{p}=.361) or disclosure-based trust with colleagues (\textit{p}=.929), which means we found no support for H3 or H4. 

\begin{table}

    \centering
    
    \begin{threeparttable}[b]
     
     \caption{Regression results (Model 1)}
     \begin{tabular}{|l|c|}

        \multicolumn{2}{c}{\textit{Spreadsheet knowledge sharing intention}} \\

        \hline

        \textbf{Variables} &  \textbf{Standardised coefficients}\\
        \hline

        Codification effort & -.295**\tnote{}\\

        Reliance-based trust in colleagues & -.085\\

        Disclosure-based trust in colleagues & -.008\\

        Software self-efficacy & .204*\\
        
        Reputational gains & .210*\\

        & \\

        Constant & 41.533\\

        $R^2$ & .270\\
        
        Adjusted $R^2$ & .231\\
        
        F & 6.948***\\

        \hline
     
     \end{tabular}
    
     \label{tab:regression_results}

     \begin{tablenotes}

     \item[] * \textit{p}$<$.05, ** \textit{p}$<$.01, ***\textit{p}$<$.001

     \end{tablenotes}
     
    \end{threeparttable}

\end{table} 

\subsection{Testing control variables}
Control variables were incorporated to test the robustness of the model. 
We evaluated the potential effects of individual characteristics such as gender \cite{beckwith2006, chintakovid2009}, occupation \cite{lawson2009}, proportion of remote work \cite{yang2021, teodorovicz2022}, and participants' self-reported spreadsheet expertise measured by their familiarity with a range of spreadsheet features. With the exception of spreadsheet expertise, none of the other control variables account for the variance in KS intention that was not already explained by the five existing predictor variables. In the model where spreadsheet expertise was included, the significant effects of software self-efficacy and reputational gains on KSI disappeared. However, expertise itself had no significant relationship with KSI, which suggests that it serves more as a mediator on the self-efficacy and reputational gains variable, rather than directly relating to KSI itself. 

\subsection{RQ2: Spreadsheet self-efficacy}
We then examined self-efficacy to test the predictions of H1b. A paired samples t-test was conducted to measure the average CSSE and GSE scores (from 1 to 7) across all participants, where a higher mean suggests a higher sense of self-efficacy and a lower mean suggests a lower sense of self-efficacy. Table \ref{tab: mean_SE} shows that, as predicted, participants' mean rating for CSSE was significantly higher than the mean rating for GSE. Even when participants were split according to their job functions into administration (\textit{n}=41) and financial (\textit{n}=50) users (excluding participants who defined their job function as `Other'), surprisingly, the difference between CSSE and GSE could still be observed in both user groups. Furthermore, the difference in self-efficacy was not significantly different between finance and administration users (\textit{t}(89)=1.59, \textit{p}=.116, \textit{d}=0.33), which suggests that, regardless of occupation, participants did not translate their confidence in spreadsheet proficiency in a work-context into a general sense of confidence of their spreadsheet proficiency. 

\begin{table}

    \centering

    \caption{Differences in mean values of general (GSE) and context-specific self-efficacy (CSSE)}

    \begin{tabular}{|p{2cm}|c|c|c|c|c|}

        \hline

         \multirow{2}{*}{\textbf{User group}} & \multicolumn{2}{c|}{\textbf{CSSE}} & \multicolumn{2}{c|}{\textbf{GSE}} & \multirow{2}{*}{\textbf{Difference in SE}} \\ 
         
         \cline{2-5}
         
         \multicolumn{1}{|c|}{} & \textit{Mean} & \textit{SD} & \textit{Mean} & \textit{SD} & \multicolumn{1}{c|}{}\\

         \hline
         
         \textbf{All} & 5.35 & 1.05 & 4.07 & 1.06 & 1.28 \\


         \textbf{Administration} & 5.16 & 1.05 & 3.66 & 1.10 & 1.50 \\


         \textbf{Finance} & 5.60 & 1.04 & 4.49 & 0.87 & 1.11 \\

         \hline

    \end{tabular}

    \label{tab: mean_SE}

\end{table}

\section{Discussion}
Despite the widespread prevalence of increasingly sophisticated tools for building formal educational resources, informal knowledge sharing remains indispensable in fostering user awareness of the functionalities and potential of these tools \cite{sarkar2023}. This study provides quantitative evidence of the extent in which personal and social variables influence spreadsheet knowledge sharing, highlighting their relevance alongside established software-related variables of interest, such as codification effort. In the following sections, we break down the key findings of our study, and explore their implications for design.

\subsection{RQ1: Personal, social, and software factors in knowledge sharing}
Overall, we found that individuals with higher levels of self-efficacy, or a belief that sharing results in reputational gains, showed greater KS intention. While the coefficients observed in the data were somewhat weak, the individual effects of each construct was statistically significant. These findings serve to extend and to provide preliminary confirmation of phenomenon previously only described through qualitative research. Previous research shows that self-efficacy predicts users' preferences for helpful `tinkering' behaviours \cite{beckwith2006, wiedenbeck2004}. Our study suggests self-efficacy also has positive implications for engagement with KS practices. Additionally, our findings provides empirical evidence to highlight that demonstrations of spreadsheet proficiency (e.g. via KS) can be associated with increased social status, which can have notable professional impacts \cite{sarkar2023, sarkar2018}. It may also explain why users appear concerned with activities such as `cleaning' and curating the presentation of their spreadsheets when sharing with others \cite{srinivasaragavan2021, chalhoub2022} -- clear presentations may be implicitly associated with a greater image of professionalism. 

In addition, our study demonstrates quantitatively that higher levels of effort associated with codifying, documenting, and communicating spreadsheet knowledge can act as a deterrent to KS intention. This provides additional insights from existing literature, which has been limited to studying instances where sharing has already occurred despite these perceived problems \cite{hendry1993, srinivasaragavan2021, chalhoub2022}. Indeed, high levels of perceived codification effort may contribute significantly to the organizational knowledge losses associated with legacy spreadsheets noted by Smith et al. \cite{smith2017}. Spreadsheets, often developed ad-hoc for specific tasks, can evolve into `templates' for long-term use despite not initially being designed for this purpose \cite{hermans2011a}, complicating the documentation process and therefore increasing the likelihood of such activities being neglected and knowledge lost \cite{smith2017, horvath2023}.

However, we did not identify a significant relationship between trust in colleagues and knowledge sharing as predicted. This was somewhat unexpected considering the results of past research \cite{giannisakis2022, wickramasinghe2012, lee2010}. Our results may be due to three factors: 1) modifying BTI items for the spreadsheet context may have affected its validity, 2) trust may not have been the most appropriate construct to capture users’ social concerns in automated knowledge sharing compared to constructs such as loss of privacy \cite{murphy-hill2015, giannisakis2022}, and 3) users' decisions to share spreadsheet knowledge may be more influenced by potential performance outcomes (e.g. helping other users so they can better carry out their respective responsibilities) \cite{srinivasaragavan2021, brown2020a} than considerations of interpersonal relationships. 

\subsection{RQ2: The paradox of self-efficacy}
We found that, despite overall demonstrating high self-efficacy when reflecting on their spreadsheet proficiency in their jobs, participants -- regardless of their occupation and objective measures of expertise --  reported low self-efficacy in their general spreadsheet expertise. The identified discrepancy in self-efficacy beliefs is potentially problematic because: 1) a conservative estimate in either CSSE or GSE may pose as a barrier to KS intention, and 2) the findings imply that users' evaluation of general software proficiency are formulated independent of their work-based performance, when the latter is likely more relevant for evaluating whether an individual is suited to KS in the workplace. This could lead to experienced individuals missing opportunities for knowledge sharing due to a false perceptions of their own qualifications. Furthermore, since expertise was found to partially explain the effects of self-efficacy on KSI, but did not itself affect KSI, this suggests that future spreadsheet designs should focus on supporting users' sense of self-efficacy rather than focusing explicitly on expertise. 

\subsection{Implications for design}
Crowdsourcing techniques from software user communities have been applied in the past to automate the KS process while simultaneously allowing learners access to a wider pool of information \cite{matejka2009, lafreniere2013a}. Applied on a more local scale (e.g. within a department or team) \cite{giannisakis2022, bateman2013}, these techniques could help to extract more contextually relevant information from each member to support the overall development of the team. However, as our study demonstrates, motivating engagement and acceptance of these systems will require sensitivity to sharers' self-efficacy and the impact such automated extraction may have on their control over their own image to others. 

Artificial Intelligence (AI) may provide a useful starting point for mediating between these varying user needs. AI's ability to summarise and provide interpretations of complex information can make the process of capturing and explaining tacit, contextual information easier for users - especially for written communication (e.g. emails, Teams channels), which our study demonstrates is a key method of KS. For example, one might envision an AI chatbot which is embedded in the spreadsheet interface, which could provide interactive explanations of the spreadsheet's editing history and the changes which have been implemented by a particular author. Alternatively, an AI assistant could be embedded in users' communication channels and help knowledge sharers evaluate the possible task complexity, estimated time, and importance of a task associated with addressing a written help-seeking request. This can support users in better evaluating how and to whom they dedicate time for, giving them greater control and decision-making with managing their reputation during KS and their ability to address others' informational needs. 

Beyond designing to support the process of KS itself, helping users build spreadsheet self-efficacy can also encourage greater engagement with KS activities in general. One way of doing so may be to implement feedback systems in the user interface. For example, highlighting the most commonly completed tasks users perform and the features frequently used in these tasks (such as in task-based user interfaces \cite{lafreniere2014}). This could help make users' generally high sense of context-specific self-efficacy more salient, allowing themselves to identify as experts for particular task types in their domain, build greater confidence, and thus enhance the likelihood of KS. Indeed, other forms of feedback systems have been shown to promote positive behaviour changes as users align their practices according to the values and priorities of the designed feedback systems (e.g \cite{malacria2013, giannisakis2022, bateman2013}.) 

\subsection{Limitations}
There are a few limitations with our study. First, the self-report nature of survey methods means there is a risk of social desirability bias in reporting. It also meant we were limited to knowledge sharers' perspectives, and could not consider the needs of knowledge recipients and their receptiveness to KS activities and recommendations \cite{brown2017}. Future research should employ observational methods or experimental paradigms to rectify this. Secondly, as our study is an early investigation into spreadsheet-specific knowledge sharing, we had to adapt existing scales. Further investigation is needed to empirically validate and confirm the robustness of our findings, such as factor analysis. Finally, we defined KS as a broad cluster of sharing activities as is the practice in knowledge management literature (e.g. \cite{lin2007}), involving both community-facing interactions (e.g. using forums) and private, one-on-one interactions, but different factors may have differing levels of relevance depending on whether sharing is public or private. Our suggested design implications for supporting self-efficacy could potentially have implications for both community-facing and private KS. However, we acknowledge that the current study does not provide enough evidence of the variables which differentiate between these types of sharing, and that the difference in KS contexts should be considered more explicitly in future research. 

\section{Conclusion}
Knowledge sharing interactions can provide a key avenue for end-user programmers to discover new capabilities in feature-rich programming environments. Our study used a multiple regressions analysis to identify that higher rates of self-efficacy and perceived reputational gains from sharing predicted greater KS intention, while higher perception of codification effort predicted lower KS intention. We reflect on the fact that participants tended to report lower general spreadsheet self-efficacy, despite also reporting high spreadsheet self-efficacy in a job-related context. We highlight how these beliefs may lead to experienced individuals unnecessarily refraining from sharing. We therefore suggest that, beyond implementing tools to automate the process of KS, a greater focus on social incentives and self-efficacy is necessary to ensure users are motivated to use such tools and engage in knowledge sharing to begin with.  

\section*{Acknowledgment}
We are grateful to our participants for their time. Nancy is supported by the Engineering and Physical Sciences Research Council [Grant number: EP/W522077/1] and Microsoft Research through its EMEA PhD Scholarship Programme.

The authors contributed the following roles in this paper. Qing (Nancy) Xia: Conceptualization, Methodology, Investigation, Formal analysis, Writing - Original Draft, Project administration. Advait Sarkar: Conceptualization, Supervision, Writing - Review and Editing. Duncan Brumby: Supervision, Writing - Review and Editing. Anna Cox: Supervision, Writing - Review and Editing. 


\appendices

\section{Participant demographics}
\label{sec:appendix_A_participant_demog}

\begin{table}[h!]

    \centering

    \begin{tabular}{|c|c|}

        \hline

        \textbf{Finance-related job functions} & \textbf{Number of participants} \\

        \hline

        Finance and accounting & 47 \\

        Banking & 3 \\

        \hline

        \textbf{Administrative job functions} & \textbf{Number of participants} \\

        \hline

        Administration & 13 \\

        Sales and distribution & 14 \\

        Marketing & 7 \\

        Human resources & 6 \\

        Other & 9 \\

        \hline

    \end{tabular}

\end{table}

\section{Items for measuring context-specific (CS) and general (G) software self-efficacy}
\label{sec:appendix_B_SE_items}

\begin{table}[h!]


    \centering

    \begin{threeparttable}[b]

        \begin{tabular}{|c|p{0.6\columnwidth}|}

        \hline

        \textbf{Item} & \textbf{Statement} \\

        \hline
    
        SE-1 (CS) & \textit{I lack the capability to effectively use spreadsheets in my job}\\

        SE-2 (CS) & I know enough about spreadsheets to get my job done\\

        SE-3 (CS) & I am confident I could complete my job well using spreadsheets even if I had no colleagues to speak to \\

        SE-4 (CS) & I am confident I could complete my job well using spreadsheets even if I did not have access to the internet\\

        SE-5 (CS) & \textit{In general, I find it challenging to obtain outcomes that are important to me when using spreadsheets in my work}\\

        \hline

        SE-6 (G) & \textit{I think my ability to use Excel can be improved substantially}\\

        SE-7 (G) & I use spreadsheets whenever I can\\

        SE-8 (G) & I have mastered Excel use\\

        SE-9 (G) & \textit{I am probably less competent than the average spreadsheet user}\\

        \hline
        
        \end{tabular}

        \label{tab:SE_scale}

        \begin{tablenotes}
            \item[] \textit{Italics indicates item was reverse coded}
        \end{tablenotes}
        
    \end{threeparttable}

   
\end{table}

\section{Full survey}
\label{sec:appendix_C_full_survey}

View the full survey by \href{https://liveuclac-my.sharepoint.com/:b:/g/personal/zcjtqxi_ucl_ac_uk/EWL9WZ5yMKRFlMhY-oNT8TkBQWQbNmiX_Ffjs78N1oIQ6Q?e=jkFnVy} {clicking here}.

\vspace{3cm}
\section{Reliability of survey scales}
\label{sec:appendix_D_scale_reliability}

\begin{table}[h!]
    \centering
    
    \begin{tabular}{|c|c|}

    \hline

    \textbf{Scale} &  \textbf{Cronbach's alpha}\\

    \hline
    
    Knowledge sharing intention & .690\\

    Codification effort & .770 \\
    
    Reliance-based trust & .867 \\

    Disclosure-based trust & .859 \\

    Spreadsheet self-efficacy & .830 \\

    Reputational gains & .816 \\

    \hline
    
    \end{tabular}
    
    \label{tab:cronbach_alpha}
\end{table}

\FloatBarrier 

\section{Frequency of common knowledge sharing practices}
\label{sec:appendix_E_KS_practices}

\begin{table}[h!]
    \centering
    
    \begin{tabular}{|p{5cm}|c|}

    \hline
    
    \textbf{Knowledge sharing practice} &  \textbf{Percentage} \\

    \hline
    
    By sending messages in communication channels (e.g. Teams, email) & 63\% \\
 
    By having informal conversations or chats with others & 62\% \\

    By annotating or commenting directly on a spreadsheet & 34\% \\

    Through scheduled meetings & 32\% \\

    By sharing resources or links to resources with others & 24\% \\

    By writing documentation outside of the spreadsheet (e.g. Word, Powerpoint) & 23\% \\

    Through impromptu demonstrations & 23\% \\

    By posting on channels or forums which can be viewed by your community (e.g. your team, department, or organisation) & 9\% \\

    Other & 1\% \\

    \hline
    \end{tabular}
    
    \label{tab:KS_practices}
\end{table}

\FloatBarrier

\bibliography{S2_Ref_Short}
\bibliographystyle{ieeetr}

\end{document}